# Controlled collision of drops in extensional flow using a six-port microfluidic device

Aysan Razzaghi[1] and Arun Ramachandran[1*]

[1]Faculty of Applied Science & Engineering, University of Toronto, 200 College Street,

Toronto, ON M5S 3E5, Canada

* Corresponding author: arun.ramchandran@utoronto.ca

Collision of two dispersed drops in the matrix of suspending liquid is the first step toward coalescence. However, to quantify the rate of coalescence, the configuration of the collision should be definable and the force that induces the collision should be measurable. We present a strategy to use the hydrodynamic force in a six-port microfluidic channel to steer two drops towards collision in the extensional flow. By implementing the analytical solution in the control loop, the flow rates that are required to steer the drops toward their respective target points can be determined using a single control parameter. This parameter, $\chi_*$, is a dimensionless time scale that can manipulate the drops in one of the two manners: 1) by engaging all six ports to create a flow field with two stagnation points ($\chi_* \ll 1$), or 2) by deactivating some of the ports and creating a linear extensional flow through the remaining active ports ($\chi_* \gg 1$). We determine specific orientations that are more suitable for the collision of Hele-Shaw drops in the six-port microfluidic channel. Based on the above strategy, we design and perform controlled head-on and glancing collisions for ~100 μm radius Hele-Shaw perfluorodecalin drops in silicone oil. The analytical solution of the flow field accounting for the perturbation of the flow due to the hydrodynamic interactions between the Hele-Shaw drops was developed using the conformal mapping technique. Coalescence time between two Hele-Shaw drops undergoing a head-on collision in a dimpled mode was found to be independent of the strain rate of the hydrodynamic flow.

# 1. Introduction

Microfluidics has been integrated with external force fields such as magnetic, electric, acoustic, and optical fields to perform manipulation, deformation, adhesion and separation of soft particles and biological cells[1–8]. Unfortunately, aside from the burden of setting up an external field in the experiments, the use of these external fields is subjected to limitations on particle and material type. For instance, the use of magnetic controllers is only limited to magnetic particles[5]. Optical forces can cause heating issues which may alter the local viscosity[9]. Also, the electric and the acoustic forces exert an inordinate stress that can damage the delicate biological cells[10]. These problems associated with external fields prompt the question - can the flow field naturally available on the microfluidic platform – the hydrodynamic field – be used to manipulate particles? The answer to this question, as has been shown in several recent studies[11–20], is an overwhelming 'yes'. In these studies, which are inspired by the four-roll mill[11,21,12], isolation and manipulation of a single micron size particle/drop was demonstrated. The trapped particle/drop experienced a measurable hydrodynamic force from the flow by showing deformation, which were used to assess the rheological properties of the interface such as interfacial tension[13,19,20,22–24]. Two drop coalescence experiments were also performed by Leal and co-workers[25–30]. But Shenoy et al.[31], inspired by Schneider et al.[16], recently took this to the microfluidic level by confining and manipulating two particles with six, evenly-distributed side channels along a hexagonal microfluidic chamber[31,32]. Model Predictive Control (MPC) was implemented as a control code to enhance the stability of control. The flow-induced interaction and adhesion of two suspended particles from 0.1 µm Brownian particles to 10 µm non-colloidal particles was demonstrated using such device [31,32]. We[33] and others[34] have adapted this device to the flow of drops. The analytical[26,33] code, in which the streamline velocity in the absence of drops is assumed to be proportional to the current and target points, and MPC code[34], exerted control over the motion of the drops in the precise manner, and collision leading to coalescence was also observed. The systems examined were light mineral oil as the suspending medium and water as the drop medium, with Span 80 in the oil phase as surfactant[33,34].

In the paper, we introduce two ideas to the manipulation and control of a pair of drops. First, the analytical code, which functions on matrix inversion to yield flow rates[12], has received little attention. The analytical code is deemed impractical for experiments because of unrealistically large flow rates associated with the close encounter of two particles. Instead, the MPC code is used, which optimises and moderates the fluxes to create the flow. However, as we shall show in the next section, the analytical code uses a single parameter - the flow rate through one of the ports – to determine the flow rates through the remaining ports. Using a single dimensionless parameter $\chi^*$ based on this flow rate, we are able to send the drops to their respective target points either by placing the stagnation points near the drops involving all six of the ports ($\chi^* \ll 1$), or by deactivating one port so that we are reduced to a five port flow ($\chi^* \gg 1$). Hence, we can trap the drops in the limit $\chi^* \ll 1$, and perform head-on or glancing collision created by the linear extensional flow in the limit $\chi^* \gg 1$. It is similar in concept to the article [22,23] above, but much simpler in execution by the switch of a button. The advantage of this method is that drops always intend to move along the straight line connecting the initial and target positions at a user-prescribed speed. Using this idea, we outline the manner in which head-on and glancing collision can be done. Second, we use a simple theory to achieve a precise control over the soft Hele-Shaw particles by using a first-order correction to Stokes flow without compromising on the speed of control. When a Hele-Shaw drop moves with a certain velocity in an extensional flow, the far field disturbance decays cubically with distance. So, if two drops are separated only by a few drop radii, hydrodynamic interactions by the method of reflections become important, and control strategies based on point particle solutions can fail[35]. In particular, we perform a head-on collision between similarly-sized Hele-Shaw drops at a variety of capillary numbers in the dimpled state, and generate a capillary number coalescence time curve. Interestingly, in the dimpled regime, the coalescence time is independent of the capillary number! In a microfluidic two-phase segmented flow, the channel-spanning drops are more likely to be generated in microfluidic devices[36]. Thus, there is interest in generating coalescence data of Hele-Shaw drops.

The article has the following parts. In Sec. 2, the new control algorithm and the creation of head-on and glancing collisions are explained. Sec. 3 deals with the experimental methods and procedure. Sec. 4 discusses the results of the control algorithm, and also presents experimental results of the drainage time as a function of capillary number. There is a discussion of analytical method and other details in Sec. 5. Sec. 6 ends with the conclusions and future work.

## 2. The control algorithm

### 2.1. Principles of the algorithm

The six-port Microfluidic Extensional Flow Device (MEFD) (Figure 1a) consists of a circular main channel of radius $R$ and depth $H$, with six evenly distributed side channels to flow the fluid in and out, thus creating a steady flow. The analytical solution of the flow field in depth-averaged co-ordinates, $\mathbf{v}^\infty(\mathbf{x}, \mathbf{Q}/H)$, is already known for point objects inside the six-port MEFD[16,31,33] and is determined by solving the Laplace equation for the streamfunction $\psi^\infty$, with appropriate boundary conditions at the six circular arcs of the boundary based on the flow rate vector $\mathbf{Q} = [Q_1, Q_2, ..., Q_6]$ (Figure 1b). The velocity field is assumed to be steady and instantaneously responsive to changes in the flow rate. Point particles are simply advanced in time by the velocity field, $\mathbf{v}^\infty = \left[\frac{\partial \psi^\infty}{\partial y}, -\frac{\partial \psi^\infty}{\partial x}\right]$, predicted by this solution. Consider, for example, two point particles, A and B, at positions $\mathbf{x}_A$ and $\mathbf{x}_B$ in the main channel's circle. The flow rates in the side channels (**Q**) should be adjusted such that the instantaneous particle velocities steer the particles from their current positions toward the target points ($\mathbf{x}_{A_T}$ and $\mathbf{x}_{B_T}$, respectively), i.e.,

$$\mathbf{v}_P^\infty(\mathbf{x}_A, \mathbf{x}_B, \mathbf{Q}/H) = \chi_A(\mathbf{x}_{A_T} - \mathbf{x}_A), \quad \text{1a}$$

$$\text{and } \mathbf{v}_P^\infty(\mathbf{x}_B, \mathbf{x}_A, \mathbf{Q}/H) = \chi_B(\mathbf{x}_{B_T} - \mathbf{x}_B), \quad \text{1b}$$

where $\chi_A$ and $\chi_B$ in s$^{-1}$ are the gain parameters. Substituting for the velocity $\mathbf{v}_P^\infty$ from Appendix A, we have

$$A\,[Q_1\ Q_2\ Q_3\ Q_4\ Q_5]^T = \begin{bmatrix} \chi_A(\mathbf{x}_{A_T} - \mathbf{x}_A) \\ \chi_B(\mathbf{x}_{B_T} - \mathbf{x}_B) \end{bmatrix}. \quad 2$$

Here, $A$, having units of 1/m$^2$, is a 4×5 matrix that gives the velocity when multiplied with the flow rates. There are six flow rates, but since the conservation of mass requires $\sum_{m=1}^{6} Q_m = 0$, there are really only five unknown flow rates. This is a vector set of four equations that is linear in the flow rate vector, **Q**. Thus, as also noted by Schneider et al.[16], Shenoy et al[31]., and Kumar et al[33]., there is an extra degree of freedom. This can be exploited to impose other constraints, but in this work, we tune the control by writing

$$A'\,[Q_2\ Q_3\ Q_4\ Q_5]^T = \chi \begin{bmatrix} \mathbf{x}_{A_T} - \mathbf{x}_A \\ \mathbf{x}_{B_T} - \mathbf{x}_B \end{bmatrix} + Q_1\,\mathbf{B}', \quad 3$$

where $A'$ is a 4×4 matrix, and $B'$ is a 4×1 vector. In this work, for ease of implementation, we have taken $\chi_A = \chi_B = \chi$; an example of $\chi_A \neq \chi_B$ will be given later. Rendering the equation dimensionless, we have

$$\mathbf{A}^*\mathbf{Q}^* = \chi^* \begin{bmatrix} \mathbf{x}_{A_T}^* - \mathbf{x}_A^* \\ \mathbf{x}_{B_T}^* - \mathbf{x}_B^* \end{bmatrix} + \mathbf{B}^*, \qquad 4$$

where $\mathbf{A}^*$ and $\mathbf{B}^*$ are dimensionless versions of $\mathbf{A}'$ and $\mathbf{B}'$, respectively, $\mathbf{Q}^* = [Q_2\ Q_3\ Q_4\ Q_5]^T/Q_1$, $\mathbf{x}$ has been nondimensionalized by $R$, and $\chi^*$ is

$$\chi^* = \frac{\chi R^2 H}{Q_1}. \qquad 5$$

$\chi^*$ is the characteristic time scale, $R^2H/Q_1$, required to sweep the volume of the MEFD circle relative to the time scale of control, $1/\chi$. $\chi^*$ has a major role to play in the control process. If $\chi^* \ll 1$, the first term in the right hand-side of Eq. 4 becomes negligible and $\chi^*$ does not feature in the leading order in flow rate. In fact, for $\chi^* = 0$, as evident from Eq. 2, the flow matrix solution is a zero-eigenvalue problem. The only solution to the problem is that velocity of the particles is zero $[\mathbf{v}_P^\infty(\mathbf{x}_A, \mathbf{x}_B, R_B, \mathbf{Q}/H) = \mathbf{v}_P^\infty(\mathbf{x}_B, \mathbf{x}_A, R_A, \mathbf{Q}/H) = \mathbf{0}]$, i.e. the particles are put in their stagnation points, the point at which the flow speed is zero. Therefore, for $\chi^* \ll 1$, particles are only displaced marginally with respect to their stagnation points. For example, we often catch two particles in the configuration shown in figure 2 (the procedure for this initial trapping is described later in Section 3.1); the two particles are mirror images of each other. Now, in the simulation, each particle has a target location of the other particle ($\mathbf{x}_{A_T}^* = \mathbf{x}_B^*$, and $\mathbf{x}_{B_T}^* = \mathbf{x}_A^*$) for simplicity; the particles are chasing each other. The $Q_1$ port is always shown at 180°. When the parameter $\chi^*$ is zero, the particles occupy the stagnation points (shown with letter S), as can be seen in Figure 2(a). In Figure 2(b), for $\chi^* \ll 1$, two stagnation points form adjacent to the position of the particles, driving the particles towards each other. All six side channels are engaged in creating the flow field in Figure 2(a) and 2(b) in this configuration.

When $\chi^* \gg 1$, $\mathbf{B}^*$ becomes negligible, and the dimensionless flow rates $[Q_2\ Q_3\ Q_4\ Q_5]^T/Q_1$ scale linearly with $\chi^*$ (Eq. 4), and some of them increase dramatically with $\chi^*$. In fact, $[Q_2\ Q_3\ Q_4\ Q_5]^T$ scale linearly with the dimensional value of $\chi$; $Q_1$ is an inactive port (Eq. 3). Hence, the job of governing flow rates lies with five ports instead of six; the extra degree of freedom is lost. The loss of the $Q_1$ port happens in the manner in Figure 2. At $\chi^* = 0.1$ [Figure 2(b)], the stagnation point towards the 240° particle moves to the corner, and at $\chi^* = 0.3$ [Figure 2(c)] it merges with the boundary circle. At $\chi^* = 1$ [Figure 2(d)], the stagnation point on the circle splits into *half* stagnation points (shown with S'), as shown in the edge. At this point, flow in the two side channels, $Q_1$, due to restriction of the normalizing port, and $Q_3$, due to a symmetry requirement, become weaker. As $\chi^* \gg 1$ [Figure 2(e)-2(f)], two half stagnation points and a single full stagnation point form in the circular main channel. Two of the side channels, $Q_1$ and $Q_3$, become completely de-active, and the flow field due to the remaining ports represents a four-port flow field Note that, in the limit $\chi^* \gg 1$, with $Q_1$ not active, there are usually five ports, but there are only four ports in Figure 2(f) due to a mirror image.

The flow field in $\chi^* \gg 1$ resembles the linear extensional flow between the two particles! It is created in between the two particles, and the line between the two also represents the converging flow of the linear field. The 240° particle is pulled towards the linear field by the two half stagnation points, and the 60° particle is drawn towards the linear field by the full stagnation point. Thus, once the particles are in the respective positions (e.g. Figure 2), just by switching to $\chi^* \gg 1$, one can create a head-on collision.

Two particles can be set to advance with two different $\chi^*$ values as well. This is shown in Figure 3, with an example that mimics Figure 2, but with $\chi_A$ and $\chi_B < \chi_A$, respectively (Eq. 2). In this case, the particles behave in the way according to the table 1.

Table 1. The behavior of the particles controlled by two different $\chi^*$ values. Note that $\chi_B$ is smaller than $\chi_A$.

| Parameters | Results |
|---|---|
| $\frac{\chi_B}{\chi_A}\chi^* \ll \chi^* \ll 1$ | Both particles will remain close to their stagnation points [Figure 3(a)]. |
| $\frac{\chi_B}{\chi_A}\chi^* \ll 1 \ll \chi^*$ | Particle at 240° enters the half stagnation point zone, while particle at 60° will remain close to a stagnation point [Figure 3(b)]. |
| $1 \ll \frac{\chi_B}{\chi_A}\chi^* \ll \chi^*$ | Particle at 240° remains in the half stagnation zone, while particle at 60° shows minor displacement from its full stagnation point [Figure 3(c)]. |

The $\frac{\chi_B}{\chi_A}\chi^* \ll \chi^* \ll 1$ is similar to the $\chi^* = 0.1$ case in Figure 2(b) as it houses particles at their stagnation points. The case of interest is $\frac{\chi_B}{\chi_A}\chi^* \ll 1 \ll \chi^*$, where particle at 240° is controlled by two half stagnation points, while particle at 60° resides at full stagnation point. This way, particle at 60° can be stationary, while particle at 240° comes and performs a head-on collision. Surprisingly, the flow pattern in $1 \ll \frac{\chi_B}{\chi_A}\chi^* \ll \chi^*$ is more similar to that in $\frac{\chi_B}{\chi_A}\chi^* \ll 1 \ll \chi^*$. This implies that for the case of $\chi_A \neq \chi_B$, as long as $\chi^* \gg 1$, the flow pattern resembles Figure 3(b) and Figure 3(c). From here onwards, we shall proceed with $\chi_A = \chi_B$.

Let us now move to different orientations of the particles relative to the $Q_1$ port. We had chosen the particles at angles of 60°-240° in Figure 2. The 120°-300° orientation is identical to Figure 2, except the ports are different. The 0°-180° orientation is interesting for high $\chi^*$ [Figure 4(a)]. There are the usual five active ports, and in addition to the full stagnation point that lies off-center, another *full* stagnation point resides at the $Q_1$ port. In figure 4(b), we show the 90°-270° orientation at high $\chi^*$. Here, there are only 3 active ports, and half of the circle including port $Q_1$ is not used! Of the three ports that are used, two serve as the inlets, and one serves as the outlet. Similarly, for high $\chi^*$, there are no stagnation points in the interior of the cross-section with respect to 30°-210° or 150°-330° orientations at the radial locations in Figure 4.

## 2.2. Implementation of head-on and glancing collisions

Based on the figures above, we now define the ways of doing head-on collision. If we place the particles directed towards each other along the orientations 60°-240°, 120°-300° or 0°-180°, then, at high $\chi^*$ [Figure 5(a)], we generate a single stagnation point, and can implement a head-on collision. For convenience, 60°-240°, 120°-300° or 0°-180° orientation shall be now called 'the axis'. For example, we show the pathway in the head-on collision in the 60°-240° orientation in a video (Video V1). The 30°-210°, 90°-270° or 150°-330° configurations, called 'the co-axial' configurations, do not create a stagnation point near the center of the device when the particles are a distance apart [Figure 4(b)], but they do create a stagnation point when the particles are sufficiently close (Video V2). However, the flow is symmetric along the axis, whereas the flow is co-axis is asymmetric. Besides, the flow along the axis always has the stagnation point in the middle of the particles, while the flow along the co-axis has a stagnation point in the corner at the beginning of the simulation, and it only moves in the center towards the end of the simulation. For the collision of Hele-Shaw drops along the co-axis where two drops come into contact before the complete evolution of the flow field, the asymmetric flow may cause an uneven distribution of the hydrodynamic force at the interface of the drops, which might lead to asymmetric deformation of the film in the coalescence time. These reasons force us to use the axial configuration instead of the co-axial configuration. For glancing collisions, instead of particles targeting themselves as in a head-on configuration, we position the origin and the target positions slightly away from the axis in concert with glancing angle $\varphi$, as shown in Figure 5(b), such that the glancing collision occurs along the axis. For example, we show the pathway in the glancing collision in the 60°-240° orientation in a video (Video V3). In both head-on and glancing collisions, we determine the strain rate ($G$) using the flow rates recorded during the experiments as the positive eigenvalue of the matrix $\begin{bmatrix} \frac{\partial v_x}{\partial x} & \frac{\partial v_x}{\partial y} \\ \frac{\partial v_y}{\partial x} & \frac{\partial v_y}{\partial y} \end{bmatrix}$.

## 2.3. The scale $\tau^* = \chi t_{loop}$

There is a second dimensionless group governing the process: $\tau^* = \chi t_{loop}$. The time $t_{loop}$ is the total time takes for the performance of the hardware and software. To have effective control, the $t_{loop}$ should be shorter than the inverse of the gain parameter $1/\chi$, i.e. $\tau^* = \chi t_{loop} < 1$. This imposes an upper limit on the parameter $\chi$. In the experiments, $t_{loop}$ is 80 ms, hence $\frac{1}{\chi} > 80$ ms. Also, the value of $\chi^* = \chi R^2 H/Q_1$ can go to around 100 in the particle manipulation and collision run. Hence, it sets a limit on characteristic time scale $R^2 H/Q_1$ to be order 10 s. Based on $R = 1$ mm and $H = 100$ μm, the value of the maximum flow rate $Q_1$ that we get is order 1 mm³/min. As we shall show in Section 3, we approximately hit this mark, with a flow rate of $Q_1 = 0.92$ mm³/min.

## 2.4. The correction for Hele-Shaw drops

The presence of a Hele-Shaw particle can modify the flow field based on the point solution considerably when the evaluation point is close to the particle position. As such, we determined an analytical solution, $\mathbf{v}_P^\infty(\mathbf{x}, \mathbf{x}_P, R_P, \mathbf{Q}/H)$, to the flow field in the presence of a circular insert of radius $R_P$ located at the position $\mathbf{x}_P$, as shown in Figure 6(a). This is done by calculating a solution to the Laplace equation for $\psi_P^\infty$, but with a circle included in the domain using a conformal mapping technique (see Supplementary Information SI. A). The velocity of a particle is then $\mathbf{v}_P^\infty = \left[\frac{\partial \psi_P^\infty}{\partial y}, -\frac{\partial \psi_P^\infty}{\partial x}\right]$. To determine the first order effect of the presence of a circular particle on the flow field, in Figure 6(b-c), two particles were set to move from their initial positions ($\mathbf{x}_{AI}, \mathbf{x}_{BI}$) toward their respective targets ($\mathbf{x}_{AT}, \mathbf{x}_{BT}$). In Figure 6 (b), particles experience $\mathbf{v}_P^\infty$ due to the presence of the other particle, but they are advanced by the flow rates that are updated based on $\psi^\infty$. The initial particle separation $\delta/R$, as indicated in figure 6(b), is 0.26. The value of $\chi^*$ is much greater than 1. Estimation of the flow rates by $\psi^\infty$ for particles that experience $\mathbf{v}_P^\infty$ leads to deviation and overlap (11.2% of the radius) of the particles in the close contact. Once $\psi_P^\infty$ is used to update the required flow rates [Figure 6(c)], particles experiencing flow disturbances manage to pass by each other without deviation of their center points. Note that $\mathbf{v}_P^\infty(\mathbf{x}, \mathbf{x}_P, R_P, \mathbf{Q}/H)$ tends to $\mathbf{v}^\infty(\mathbf{x}, \mathbf{Q}/H)$ for $R_P \ll R$, hence the analytical solution is valid for point particles as well.

## 3. Experimental setup and procedure

To illustrate the implementation of the control, experiments were conducted by manipulating two Hele-Shaw perfluorodecalin (Sigma Aldrich, viscosity of 5 cP) drops in 500 cP silicone oil (Sigma Aldrich). Silicone oil was pumped into the six port MEFD through rigid PEEK tubes (0.02 inches ID). As shown in Figure 7, all six fluid reservoirs were connected to the pressure controller (Elveflow OB1, MK3) with a rise/fall response time of 20 ms. The dispersed phase liquid reservoir was also connected to a pressure controller (Marsh Bellofram). Perfluorodecalin drops were generated using a T-junction by means of step emulsification[37,38] and applying the "drop-on-demand" principles [39]. It is important to have a stagnant perfluorodecalin-silicone oil interface at the T-junction throughout the control to prevent the creation of new drops during the control process, which was achieved by a static pressure balance at the T-junction[39]. The microfluidic channel was etched on a silicon wafer and anodically bonded to a borosilicate glass. For accurate control over pancake-shaped soft particles/drops, it was essential to prevent the wetting of the glass and silicon channel walls by the perfluorodecalin drops. To achieve this, we coated[40] the microfluidic device with 5% (w/w) poly methyl pentene in hexane (Sigma Aldrich) by applying a gas-templating method[41].

Hele-Shaw drops were controlled in the channel by the correction proposed in section 2.4. To control drops, each control loop started with the acquisition of the MEFD circle image by a camera (Teledyne Lumenera infinity3S-1URM) at 60 fps. The captured image was sent to the computer for image processing and computation of the required flow rates using MATLAB® (details of the image processing as well as the schematic of the control loop are presented in the SI. B). The flow rates were converted to the pressure considering $P_m - P_0 = \mathcal{R}_m Q_m$, where $P_m$ is the pressure at each fluid reservoir, $P_0$ is the pressure at the center of the main circular channel, and $\mathcal{R}_m$ is the hydrodynamic resistance across each side channel and its associated PEEK tubing. The pressure $P_0$ was deduced from the incompressibility of the fluid flowing in the device, to be $P_0 = \frac{\sum_{m=1}^{6}\left(\frac{P_m}{\mathcal{R}_m}\right)}{\sum_{m=1}^{6}\left(\frac{1}{\mathcal{R}_m}\right)}$.

In the case of equal hydrodynamic resistances in the six side channels, $P_0$ reduces to $P_0 = \frac{1}{6}\sum_{m=1}^{6} P_m$. For a typical collision experiment where maximum value of $P_m$ is 5 psi and $P_0$ is approximately 3.4 psi, the resistance of $\mathcal{R}_m = 7 \times 10^{14}\ \frac{\text{Pa.s}}{\text{m}^3}$, yields the maximum flowrate of 0.92 mm$^3$/min.

To perform collision experiments, after two or more drops were found in the channel, two drops needed to be brought to the axial configuration (60°-240°, 120°-300° or 0°-180°) so that the head or glancing collisions could be initiated, while the rest of them were flushed away. For this, the target positions were placed along the axis, and control was performed at low $\chi^*$ or high $\chi^*$. Even if high $\chi^*$ is used, as the particles near their respective target positions, as indicated in Eq. 4, $\chi^*(\mathbf{x}_{A_T}^* - \mathbf{x}_A^*)$ and $\chi^*(\mathbf{x}_{B_T}^* - \mathbf{x}_B^*)$ was eventually small, and particles were finally placed at their stagnation points. Pressure at all the six side channels (plus the side channel connected to the dispersed phase reservoir) were set to zero, which stalled the flow and trapped drops. Thereafter, the control code indicated in Figure 5(a) for head-on collisions or Figure 5(b) for glancing collisions was carried out.

## 4. Results and discussion

### 4.1. Demonstration of control

To demonstrate the accurate manipulation of Hele-Shaw drops, first, two drops (100 and 125 μm radii in the device with depth of 100 μm) are kept at separate but constant initial position within 500 μm inter-drop distance for 100 seconds in the configuration shown on the inset of Figure 8(a). The effects of the variation of $\chi^*$ on the precision of control is also shown in Figure 8(a). Both drops experience in-phase fluctuations of position ($d$) with the variation of $\chi^*$. As they approach the targets, the prescribed velocity ($\mathbf{v}_P^\infty$) that is dictated by $\chi^*$ decreases due to a reduction in separation. Our results show that increasing $\chi^*$ as the drops approach the target, enhances the accuracy and reduces the offset. This can be seen at $t = 95$ s in Figure 8(a), where increasing the $\chi^*$ from 50 to 100 leads to a sharp reduction of $d$. In Figure 8(b), the set of drops in Figure 8(a) was steered along the edges of a 500×500 μm square (2 mm in total) (see the Video V4). As can be seen in Figure 8(b), the drops move in the direction of the straight line that connects the center of the drops to its target points. Figure 8(c-e) shows the positions of the centers of the drops during the manipulation in Figure 8(b). While drops move monotonically, there are some deviations of the center of the drops from the straight line connecting the center to the target points. However, as it is shown in the SI. C, these deviations are small compared to the size of the drops without any overshoots, which is a direct consequence of the matrix inversion control algorithm adopted in this work[32]. It takes 60 seconds for the drops to complete the entire trajectory, covering a length that is an order of magnitude larger than the size of the drops. The average velocity of Hele-Shaw drops in our experiments is 30 μm/s.

## 4.2. Demonstration of head-on and glancing collisions

Figure 9(a) and 9(b) demonstrates the collision of a pair of Hele-Shaw drops within head-on and glancing collisions, respectively, using the control strategy explained in Section 2.3 and Figure 5 (see the Supplementary Video V5 and Video V6). The value of $\chi^*$ was set to 100 in the collision experiments, so while drops are moving toward each other, a single stagnation point forms in between them (section 2.2). As the strain rate $G$ is a function of position in the correction for Hele-Shaw drops, its value is slightly different at the center of each drop. So, the average value of the $G$ was considered in analyzing the collision experiments. In a head-on collision [Figure 9(a)], once drops come into contact, a constant hydrodynamic force pushes the drops against each other indefinitely until the drops coalesce within a drainage time (*t*). The drops are not able to separate away due to the control over both the particles. We chose *t* = 0 corresponding to the moment where the distance between the two Hele-Shaw drops was $R_{p_1} + R_{p_2}$, the radii of the two drops. On the other hand, in a glancing collision [Figure 9(b)], the two drops collide and rotate until the point where the hydrodynamic force changes the sign, and the drops separate away[29].

## 4.3. Drainage time measurement in a head-on collision between drops

We have measured the hydrodynamic drainage time for two Hele-Shaw drops that come into contact through head-on collision in the extensional flow. Figure 9(c) shows the dimensionless drainage time (*tG*) variation with $Ca$ number $\left(Ca = \frac{4\mu G R_p^3}{\gamma H^2}\right)$[19], where $G$ is the strain rate of the flow, $\mu$ is the viscosity of the suspending fluid and $\gamma$ is the interfacial tension (IFT) of perfluorodecalin-silicone oil interface that is measured, using an in-house four port MEFD, to be 1.4 $\frac{mN}{m}$. The details of calculations of $\gamma$ measurements are discussed in the SI. C. The drainage time was measured from *t* = 0 until the instant two drops merge. To the best of our knowledge, this is the first time that the hydrodynamic drainage time is measured experimentally at a mobile interface and in the limits of Hele-Shaw motion for two drops. The linearity of the $tG$ - $Ca$ curve in Figure 9(c) suggests the independence of the drainage time from the strain rate in head-on collision! The theory supports this result (derivation given in SI. D) :

$$t \sim \frac{\lambda \mu R_P}{\gamma} \frac{\sqrt{H}}{\sqrt{h_f}}. \qquad 4$$

Here, $\lambda$ is the dispersed to suspending phase viscosity ratio and $h_f$ is the final height of the film. The cause for this is as follows. Continuity requires that the rate of volume decrease is equal to the drainage velocity multiplied by the cross-sectional co-ordinates. It is known theoretically that for unconfined drops, the rate of the volume decrease is the square of the film radius, but the velocity term is directly proportional to the film radius, hence the drainage decrease rate is inversely proportional to the film radius[29,42]. However, for the Hele-Shaw drops, the film radius features linearly on both sides of the continuity equation, and its effect on the strain rate vanishes[43].

## 5. Discussion

A few comments about the analytical method of collision in the six-port device are in order here. The flow rate matrix (**Q**) is rendered dimensionless using the positive flow rate at the first port ($Q_1$). This non-dimensionalization can be performed based on the flow at any of the 6 ports, and by assigning a positive or negative flow to the port in question. So, for a single value of $\chi^*$, there exist 12 different solutions to the topology. But as far the axes and co-axes are concerned, the head-on and glancing collision will appear to be the same, as shown in figure 2 and figure 4.

In order for the particles to get from one position to another, particle encounters are discouraged, as that leads to a high dissipation rate[16,31,32]. This can be done via MPC by employing $\beta$, the regularizer for flow rate[31,32]. In this study, for the initial manipulation to get particles to the axis, we use, not MPC, but our analytical algorithm at low $\chi^*$ or at high $\chi^*$, as indicated in Section 3. If the flow rates are unusually large, which are detected by the pressure drops, and if we are not concerned about the strain rate, which is valid in this case, we can always divide the flow rate by a factor to keep the flow rate in control. But after the setting of the particles on the axis is done, the idea is to create an encounter or a collision; therefore, we cannot invoke the minimum dissipation idea. Moreover, a single stagnation point is generated in the interior of the channel for head-on or and glancing collisions along the axis in the limit $\chi^* \gg 1$ using the analytical method. These are created by flow through five ports at $\chi^* \gg 1$, not six : the extra degree of freedom is lost (Section 2). If a collision to be performed at a controlled strain rate, which is modulated by the flow rate, there is only one option of flow rates along the axis according to the analytical solution, not multiple.

The half-stagnation points generated in the N-port device are of particular interest to us, in the light of the idea that 2N+1 ports are required to control N particles[16]. They occur in our four-port device as well - one can have a single stagnation point, or two half stagnation points. In the five-port device, a single stagnation point and a half stagnation point, or three half stagnation points could exist in the device. In the six-port device, two stagnation points, or one stagnation point and 2 half stagnation points, or 4 half stagnation points could persist. One can have a half stagnation point in three-port device as well, and trap a particle[16] ! Our preliminary investigations also indicate that in a five-port device, collision may also be carried out, as indicated by Shenoy et al[32]. This is left to future work.

We have used the Hele-Shaw microfluidic channel in this work. In a Hele-Shaw channel, width is greater than the depth. Hence, vorticity is negligible, and rotational flows are barely permitted; one has to resort to extensional flows only. For example, in a four-port MEFD, one needs to resort to extensional alignment of a particle; there is no rotational help to assist motion. Therefore, short term head-on collision is possible with a four-port device. On the original, computer controlled four roll mill[26], two drops were brought into head-on collision in an extensional flow, but because of the rotation possible in the four-roll mill, the drops could be kept infinitely long in the configuration. But on a microfluidic, Hele-Shaw, four-roll mill, we cannot do head-on collisions of two drops for relatively long period of time, as there is no rotational help. With a six-port MEFD, however, the two drops are controlled by separate target points (Figure 5), and therefore, the head-on collision can be performed on the two drops as long as the liquid lasts in the pressure reservoirs.

The total loop time in our experiments is approximately 80 ms with the following time breakdown:

(1) grabbing the image by camera (~17 ms)

(2) performing image processing (~45 ms),

(3) computing of the flowrates and pressures through the analytical solution (~13 ms), and

(4) communicating with pressure controllers (~5 ms).

The comparable total loop time corresponding to MPC, according to Ref 31, is 33 ms. This is obviously shorter than our 80 ms time scale, but remember that the analysis was done using Labview® , an advanced data analysis and control software. We have done the computation and analysis using an interpreted language, MATLAB® , which is considerably slower. If Labview® is used, our computations would be much faster.

## 6. Conclusions

In the current manuscript, manipulation of a pair of particles along the desired trajectories is demonstrated by employing the matrix inversion method to determine the required flow rates in a six-port MEFD. In past practices, the use of matrix inversion method to determine the flow rates has been discouraged as it leads to high dissipation rate for the trajectories that involve encounter of particles. However, this manuscript shows that matrix inversion method can effectively steer particles towards their respective targets by using a single control parameter, $\chi^*$. $\chi^*$ is the characteristic time scale required to sweep the volume of the MEFD circle relative to the time scale of control. Changing the value of $\chi^*$ from zero to its maximum, which is determined based on the loop time, leads to variation in the topology of the flow field from two stagnation point flow ($\chi^* \ll 1$) to the linear extensional flow ($\chi^* \gg 1$). The system was used to design and carry out systematic head-on and glancing collision between a pair of Hele-Shaw drops. To account for the hydrodynamic interaction between the Hele-Shaw drops, a correction was calculated by eliminating a circular drop in the six-port device by using the conformal mapping technique.

For confined drops undergoing the head-on collision, drainage time in the film were insensitive to the strain rate, whereas for unconfined drops, the drainage rates have shown a strain rate power of $1/3$[29,42]. This was explained on the basis of the continuity equation. The film radius corresponding to the dimpled film is the primary quantity which affects the capillary number, and continuity requires that the film radius be cancelled out, hence the result.

There are several implications of this work, and we name two here. The coalescence of two Hele-Shaw drops will require a much longer investigation. There are several variables to be explored (viscosity ratio, radius of the drop, channel depth and width, etc.), and these will be investigated to quantify drainage time in head-on as well as glancing collisions. Coalescence of drops of two different phases in the third suspending phase, which leads to formation of compound drops[44] (from core-shell to Janus) can be performed using the methodology explained here.

**Acknowledgements**

We acknowledge the funding support from Syncrude Canada Ltd and NSERC Collaborative and Research Development Grant CRDPJ 514675-17. A. Ramachandran acknowledges Canada Research Chair (File # 950-231567).

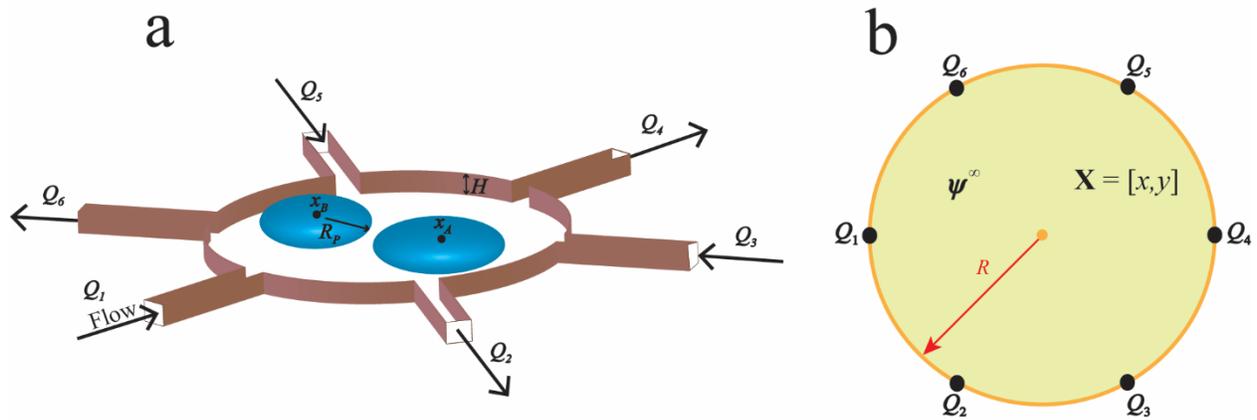

**Figure 1**. (a) Schematic of the six port MEFD with two channel-spanning soft particles, where the radius of the particles, $R_p$, is greater than the channel depth, $H$, so that the particles adopt the shape of a pancake. The flow of the fluid through the side channels leads to the creation of the flow field inside the circular main channel. (b) Geometry of the six-port MEFD circle and the flow field ($\psi^\infty$) developed for the point objects.

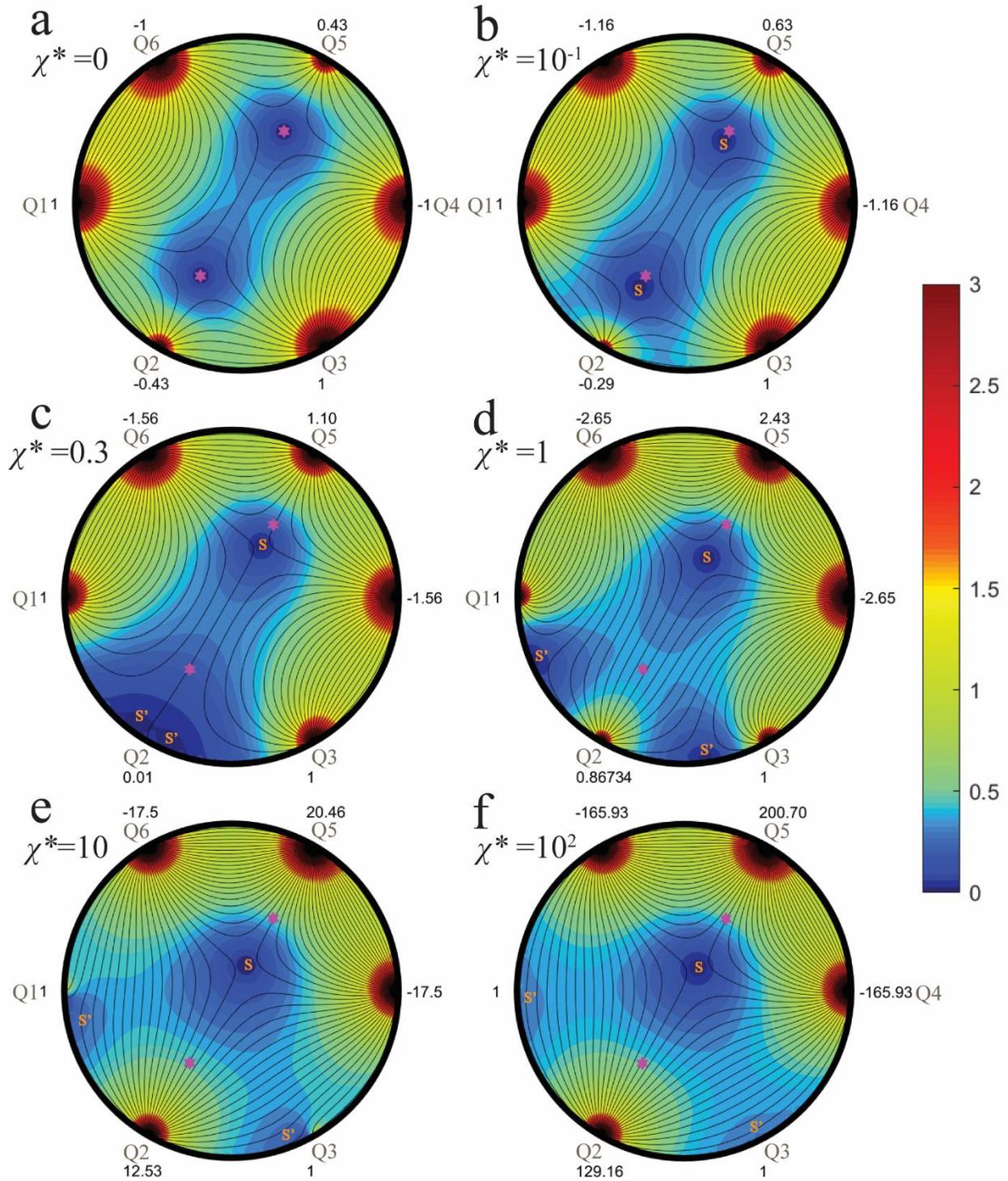

**Figure 2.** Velocity contour and streamlines for two paerticles in the 60°-240° orientation (shown in magenta hexagrams) that are set to move toward each other at a) $\chi^*=0$, b) $\chi^*=10^{-1}$, c) $\chi^*=0.3$, d) $\chi^*=1$, e) $\chi^*=10$, and f) $\chi^* = 10^2$. The stagnation points are shown with S and the half stagnation points in c-f is shown in S'.

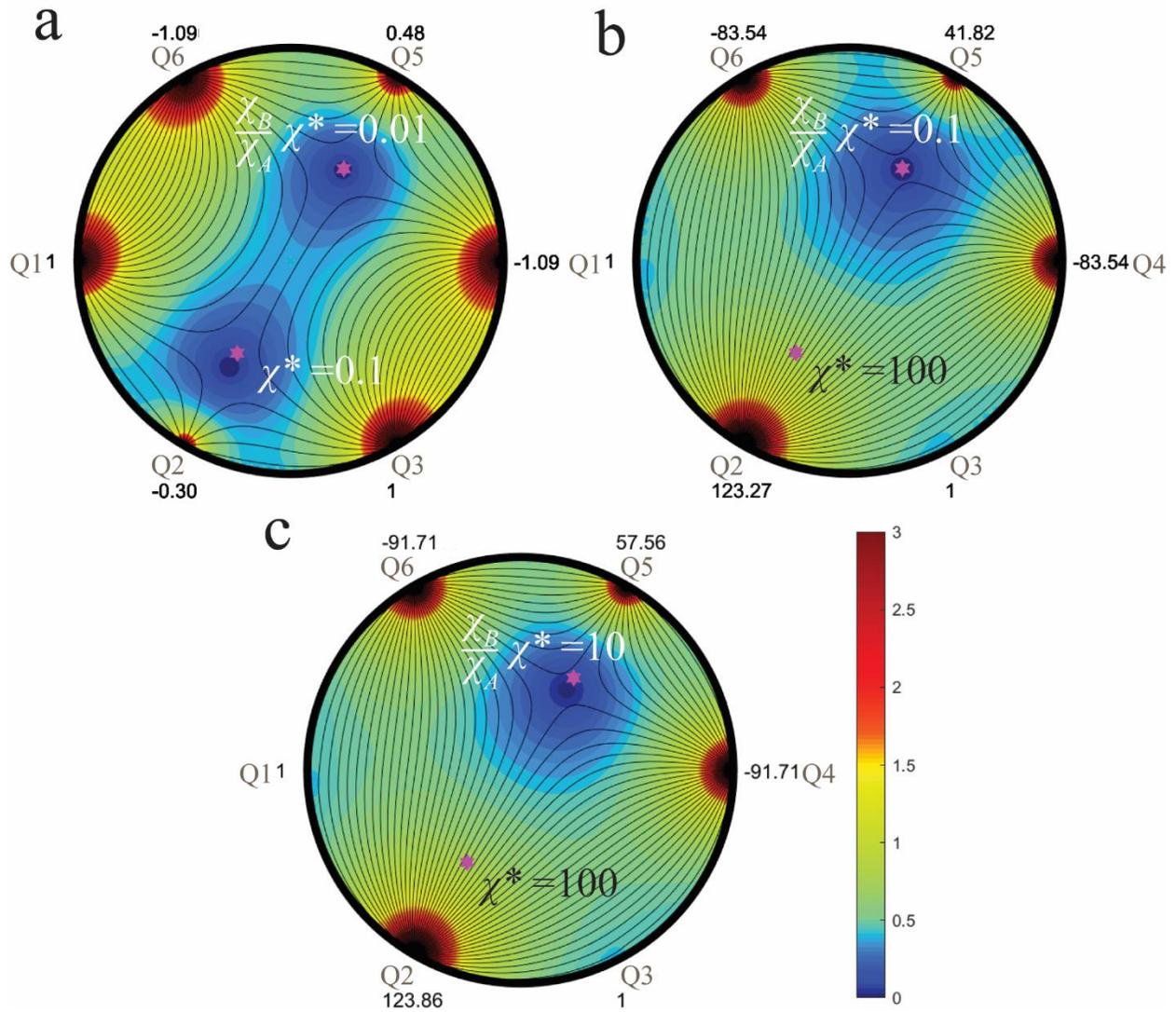

**Figure 3.** Velocity contour and streamlines for two paerticles in the 60°-240° orientation (shown in magenta hexagrams) that are set to move toward each other with different $\chi^*$ values. Assuming that $\frac{\chi_B}{\chi_A} < 1$, at a) $\chi_A^* = 0.1$ and $\chi_B^* = 0.01$, b) $\chi_A^* = 100$ and $\chi_B^* = 0.1$, and c) $\chi_A^* = 100$ and $\chi_B^* = 10$.

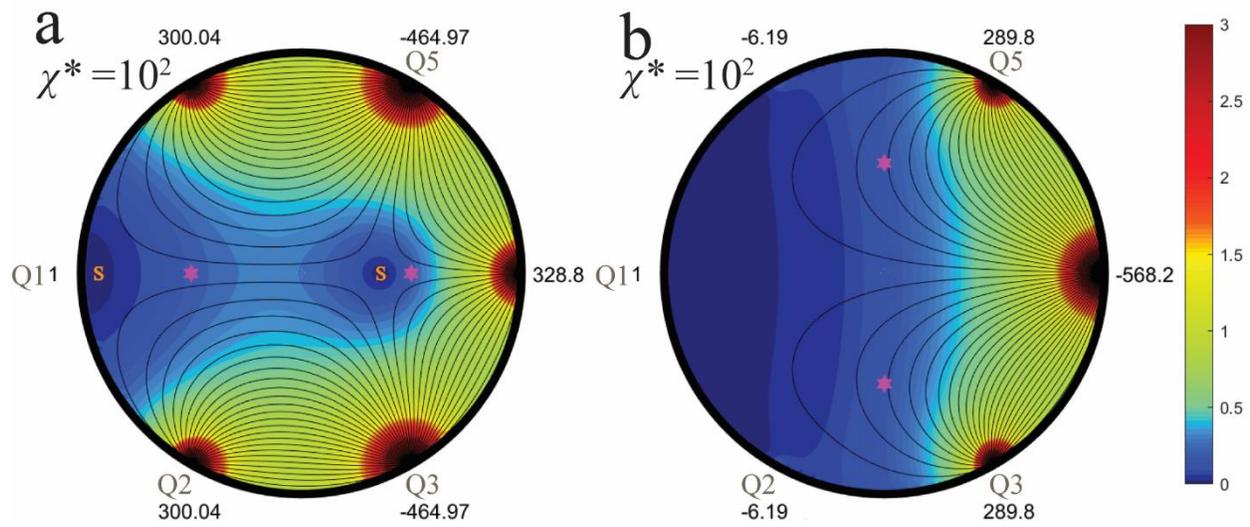

**Figure 4.** Velocity contour and streamlines for two paerticles (shown in magenta hexagrams) that are set to move toward each other with $\chi^* = 10^2$  a) in the 0°-180° orientation and b) in the 90°-270° orientation. The stagnation points are shown with S.

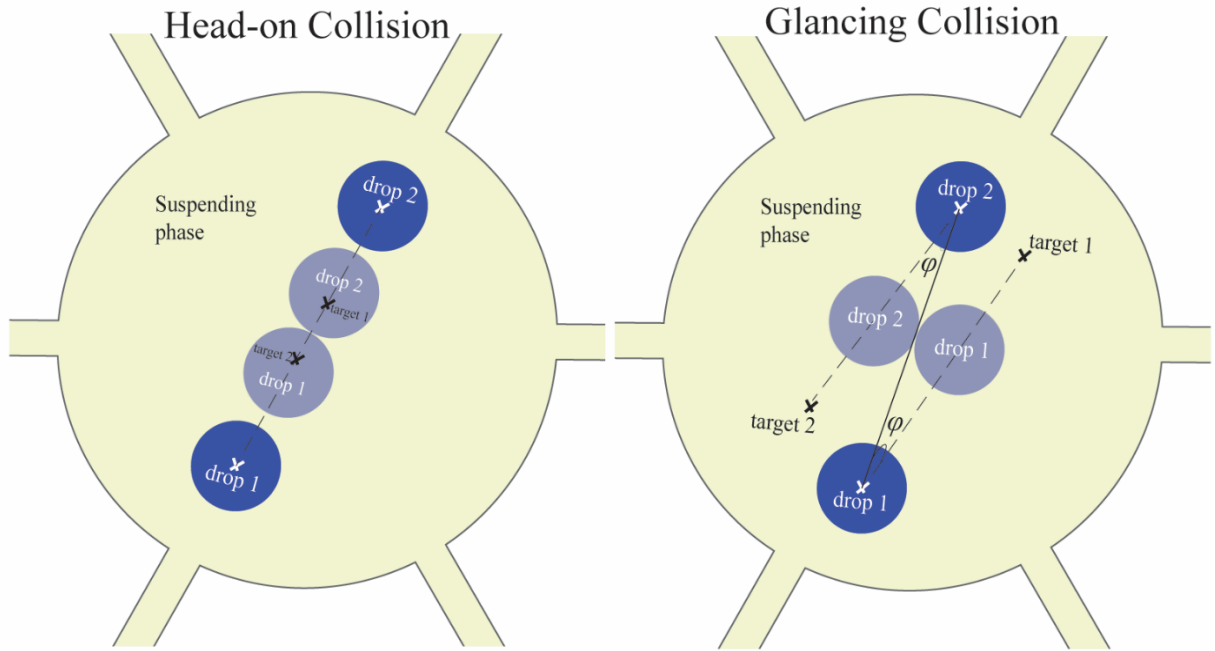

**Figure 5**. Schematic of the strategy to induce head-on and glancing collision in the six-port MEFD. For head-on collision the target points are placed on the drops' centerline. For glancing collision, targets are placed within angle $\varphi$ from the centerline.

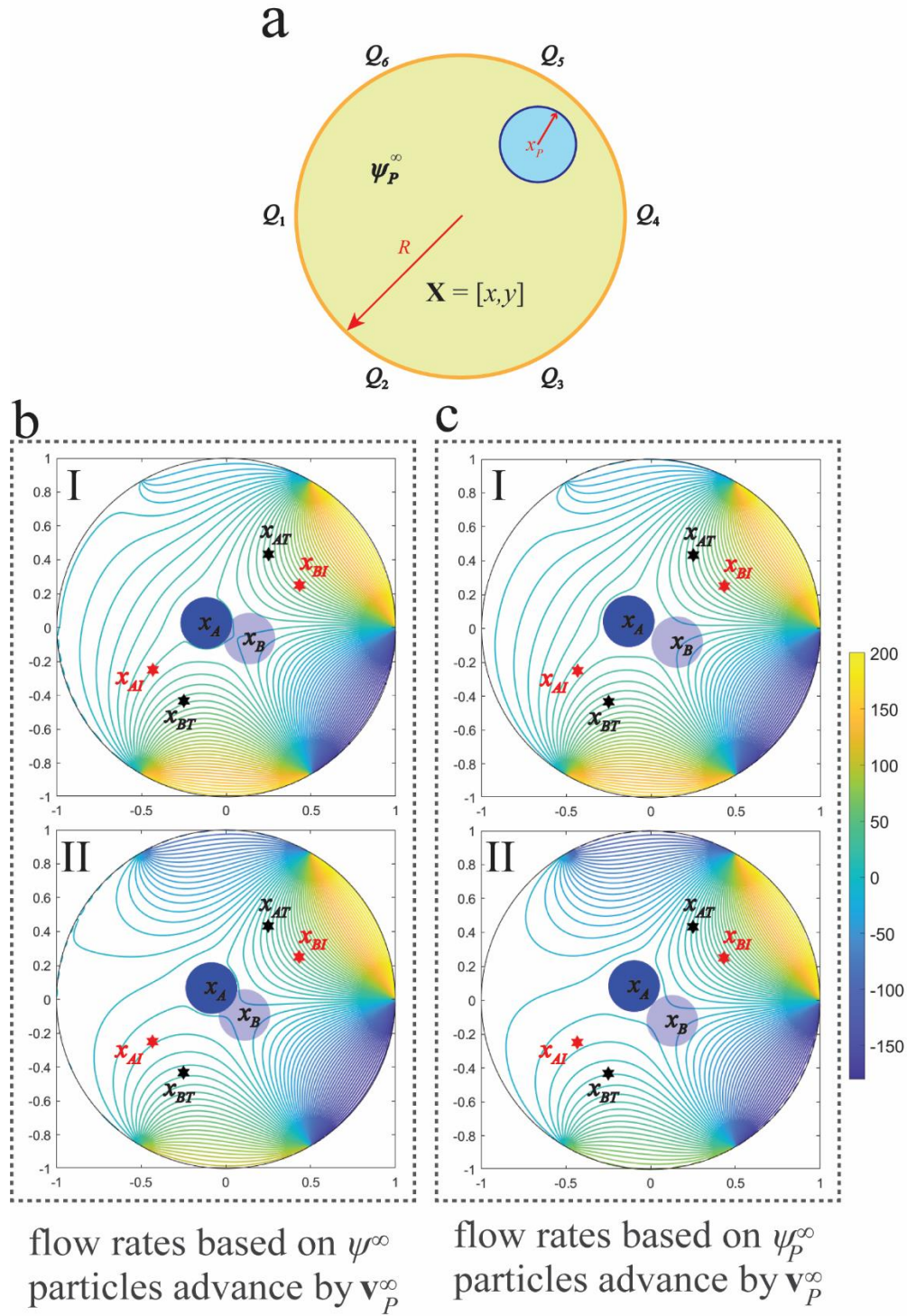

**Figure 6**. (a) Geometry of the six-port MEFD circle and the flow field ($\psi_p^\infty$) developed based on the presence of the circular insert at an arbitrary position, $x_P$. (b) The flow field evolution during the particles manipulation from initial positions ($x_{AI}$, $x_{BI}$) to target positions ($x_{AT}$, $x_{BT}$) while the particles are advanced by $\mathbf{v}_P^\infty$ and flowrates are updated

based on the $\psi^\infty$. (c) The flow field evolution during the particles manipulation from initial positions $(x_{AI}, x_{BI})$ to target positions $(x_{AT}, x_{BT})$ while the particles are advanced by $\mathbf{v}_P^\infty$ and flowrates are updated based on the $\psi_p^\infty$.

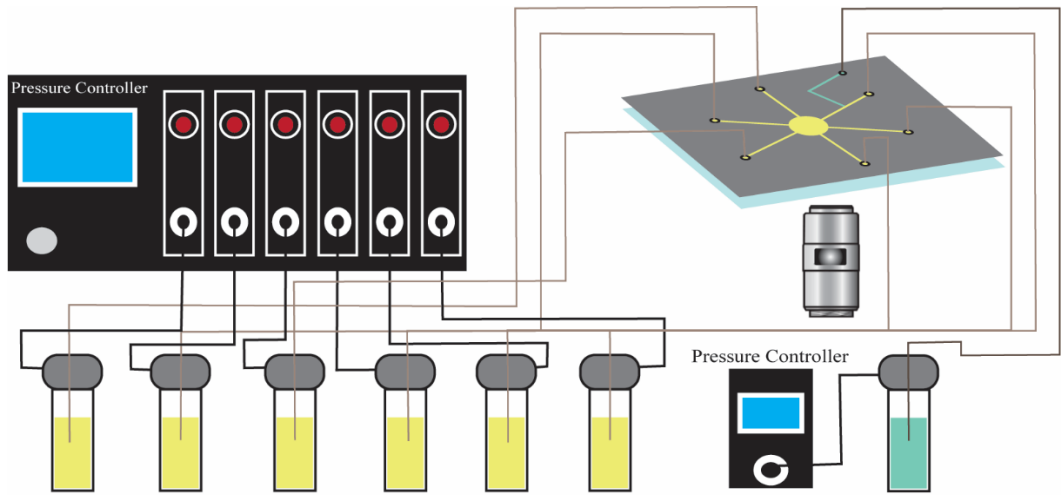

**Figure 7**. The experiment setup where the six reservoirs, each delivering fluid to one of the six-ports in the MEFD, are connected to the pressure controllers that are controlled using MATLAB®. Yellow represents the suspending/continuous liquid medium and the blue is the dispersed phase that leads to a T-junction for drop production.

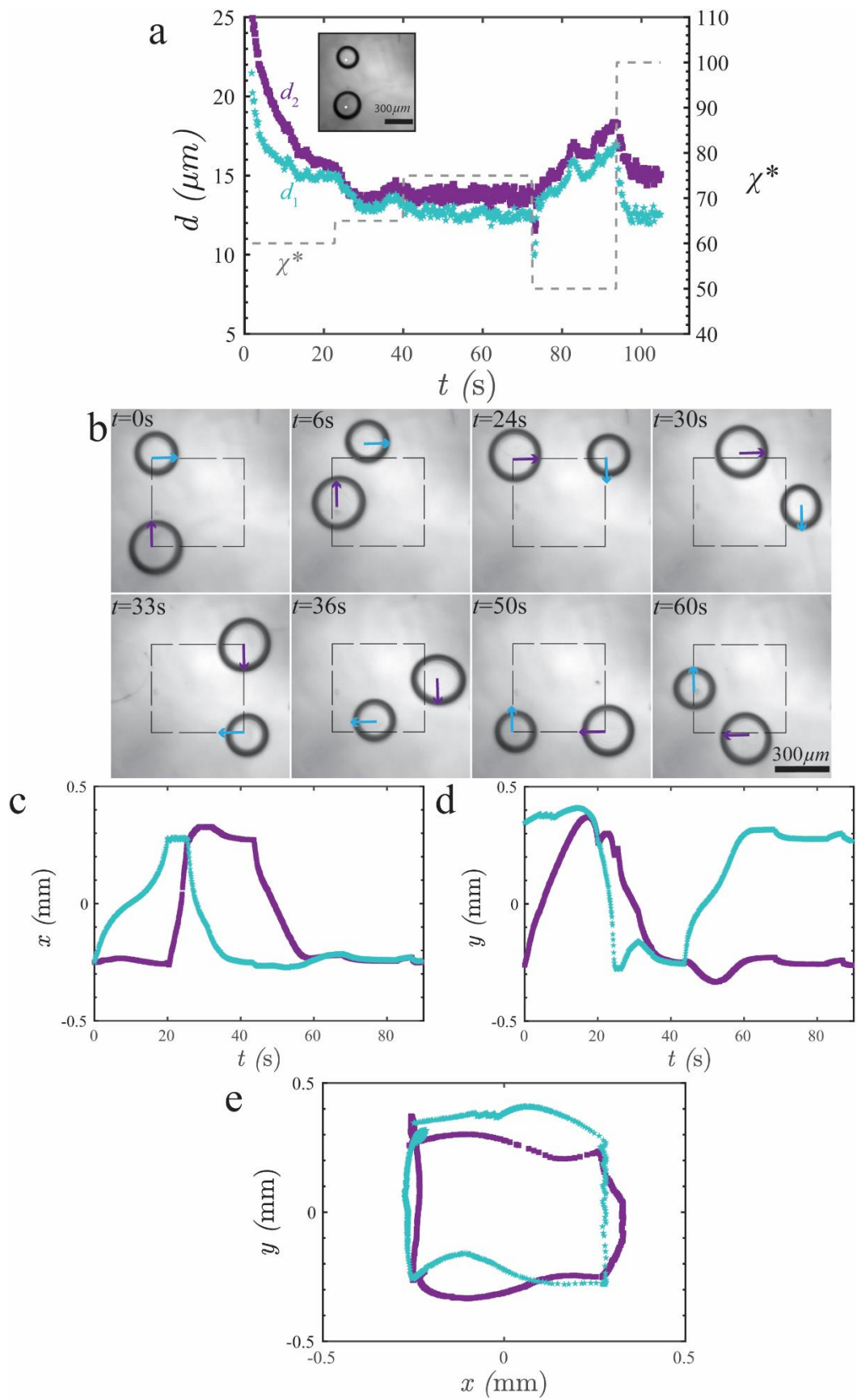

**Figure 8**. Manipulation of two Hele-Shaw perfluorodecalin drops (100 and 125 microns in radii) in silicone oil. Depth of the channel ($H$) is 100 microns. (a) Drops are held at a certain position for 100 seconds through which the distance of the drops' center from target position ($d$), shown on the left axis, fluctuates within 5 to 8% error. Values of $\chi^*$ during the manipulation (dashed gray line) is shown on the right axis. (b) Drops are steered along a predefined path. The value of the $\chi^*$ varies in the range of 100 to 180 actively, with a larger $\chi^*$ as the drops approaching the target points and a smaller $\chi^*$ while drops are far away from the targets. The total loop time for this case is 84 ms. When the two drops reach within a certain distance from the target points (20 microns in this experiment), the target points are updated, and the drops are steered toward the next set of target points. (c-d) The variation of the drops center position in x and y direction, with time, and (e) the trajectories of both the drops centers along the predefined path in b.

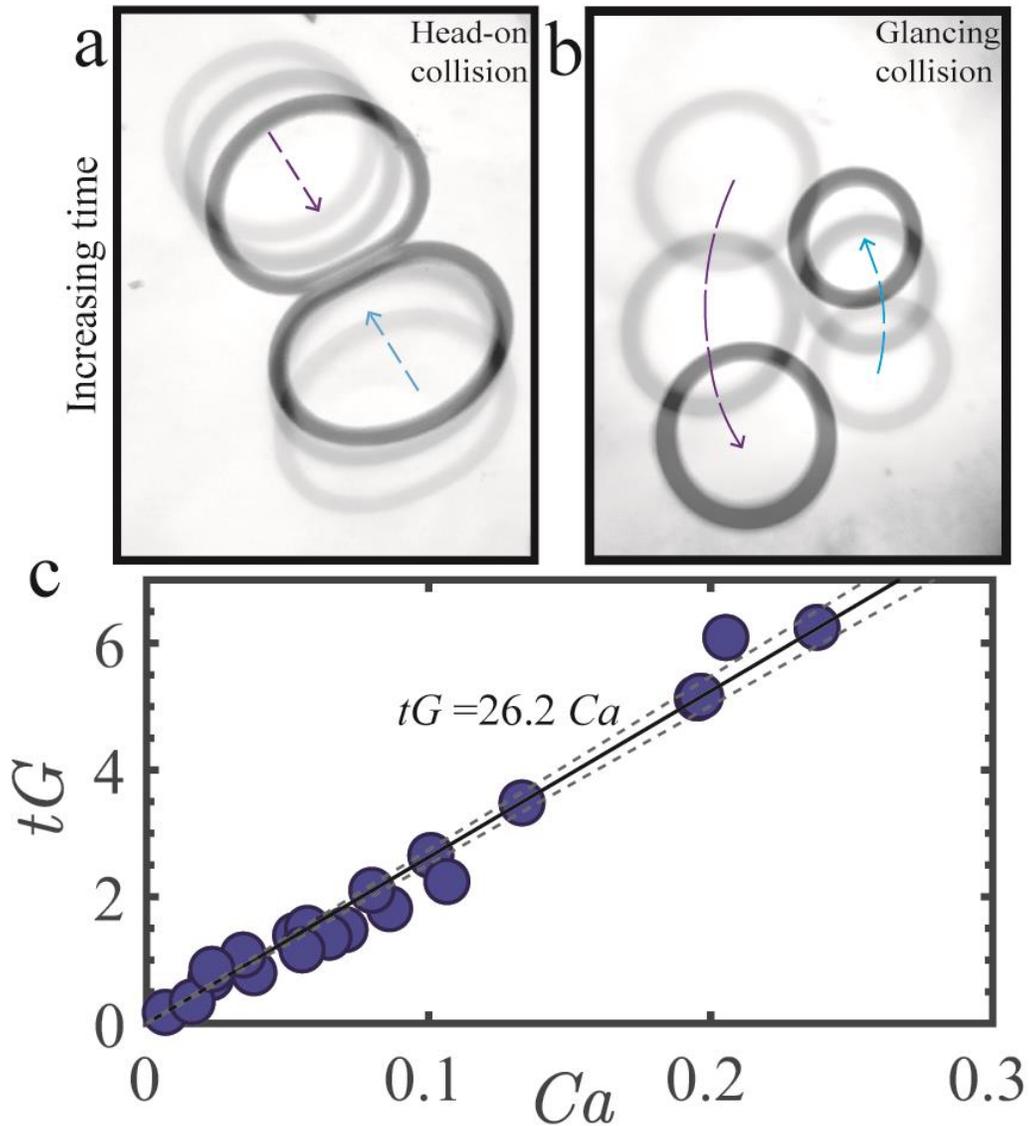

**Figure 9**. Demonstration of two Hele-Shaw perfluorodecalin drops colliding within (a) head-on and (b) glancing configurations. The time taken for glancing and head-on collisions is 13 and 6 seconds, respectively. (c) Dimensionless drainage time ($tG$) as a function of Capillary number ($Ca$) number for head-on collision of the Hele-Shaw ($R_p = 110 \pm 13\ \mu m$) perfluorodecalin drops in silicone oil, the drainage time scales as $t \sim \frac{\lambda \mu R_p}{\gamma} \frac{\sqrt{H}}{\sqrt{h_f}}$, where $\lambda$ is the viscosity ratio, $\mu$ is the viscosity of suspending phase, $\gamma$ is the interfacial tension, and $h_f$ is the final film thickness. The equation of the linear fit is shown on the graph with 95% confidence bounds (25.06,27.39) shown with dashed lines.